\documentclass[12pt]{article}
\usepackage{epsfig}

\newcommand{\beq}{\begin{equation}}
\newcommand{\eeq}{\end{equation}}

\begin{document}

\begin{center}
SOLUTIONS OF THE PION DISPERSION EQUATION IN THE MEDIUM:\\
 NEW ASPECTS\\
 V.A.Sadovnikova\\
Petersburg Nuclear Physics Institute, Gatchina, St.Petersburg
 188350, Russia, tel.812-71-46096, fax.812-71-31963,
e-mail: sadovnik@thd.pnpi.spb.ru\\
\end{center}

\begin{center}
Abstract
\end{center}
In  symmetrical nuclear matter the solutions of pion
dispersion equation are investigated in the complex plane
of the pion frequency $\omega$.
There are three well-known branches of solutions on the physical sheet
:  sound, pion and isobar, at the matter density less than the critical
one $\rho <\rho_c$. At $\rho >\rho_c$ the fourth branch appears on the
physical sheet. For this branch the condition $\omega^2_c\leq0$ (in
general case Re$\,\omega^2_c\leq0$) takes place. This points out the
instability of the ground state which is possibly related to
the pion condensation.
\vspace{1cm}

matter

\section{Introduction}
Some time ago a bright phenomenon of pion condensation
predicted in \cite{1}  attracted the common attention and was
widely investigated \cite{1}-\cite{4}. The reason for this
prediction was the fact that one of the solutions of the pion
dispersion equation in medium (let it be called $\omega_c$) turns to
zero $\omega^2_c(k)=0$ (at some momentum $k\ne0$) while
the density of medium increases \cite{1}. This meant an
appearance of excitations with zero energy in medium. In this case
the ground state should be reconstructed in correspondence with phase
transition.

One variant of reconstruction was to include  the pion
condensate (taken in one or another form) into the ground
state. This resulted in a prediction for pion condensation.
A number of efforts was devoted to the search for this
phenomenon. The result of discussions given in the book of
A.B.Migdal et. al. \cite{2} was that  the pion condensation
manifested itself weakly and probably was absent (at least not
observed) at normal density. Nevertheless, this
phenomenon could influence the equation of state at high
densities achieved in heavy ion collisions or neutron
stars.

Below, we consider in detail the solutions of pion dispersion
equation
\beq \omega^2-k^2-m^2_\pi-\Pi(\omega,k,p_F)\ =\ 0\
\eeq
in the complex plane of the pion frequency $\omega$.  In (1) $\Pi$ is
the pion polarization operator (pion self-energy) in the matter.
The consideration in the complex $\omega$-plane allows us to obtain
 additional information about  well-known solutions.

The main content is as follows.
We consider excitations with quantum numbers $0^-$ in
symmetrical nuclear matter. At the equilibrium density
there are three branches of solutions of (1). When the density
is larger than critical one ($\rho>\rho_{c}$) the fourth branch
appears on the physical sheet.

The equation (1) has logarithmic cuts on the physical sheet of
the complex $\omega$-plane, determined by the structure of the
polarization operator $\Pi$. Different branches of
solutions are considered on physical and unphysical sheets
of the complex $\omega$-plane. With the change of $k$ they move from
physical to unphysical sheet (and backward) through the cuts. Below, to
explain the obtained results, we deal with the case when isobar width
in medium is equal to that in vacuum, i.e. 115 MeV.  In this case all
solutions of (1) and certain cuts move from the real (or imaginary)
axis to the complex plane. This helps us to trace the $k$-dependence of
the solutions. For certain important cases the influence of
$\Gamma_\Delta$ on the behaviour of $\omega_i(k)$ is studied.

The well-known branches of solutions (1) are: 1) spin-isospin
sound branch $\omega_s(k)$; 2) pion branch $\omega_\pi(k)$;
3) isobar branch  $\omega_\Delta(k)$.
At $0\leq k\leq k_f$ they are located on the physical sheet (the
momentum $k_f$ is different for each branch), then with further
increase of $k$ they move to unphysical sheet across the cuts.
For these branches Re($\omega_i^2)>0$ everywhere on the physical sheet.

Our investigations show that there is the whole set
of solutions at unphysical sheets. With increasing density
certain solutions come to the physical sheet. In such a way
the fourth branch, $\omega_c(k)$, appears on the
physical sheet at $p_F\ge283$ MeV\footnote{For our
values of the parameters, which are presented below.}.
 Below $\omega_c(k)$ is referred as the condensate branch.
At $k=0$ $\omega_c(k)$ is located at the same point as
$\omega_\pi(k)$; with increasing of $k$ the branch goes onto the
unphysical sheet.  At some $k=k_1$ it appears on the physical sheet and
at $k=k_2$ leaves it for the same unphysical sheet. The values of $k_1$
and $k_2$ depend on the density.  At critical density, $\rho_c$,
corresponding to $p_F=283$ MeV ($\rho_c\simeq 1.2\rho_0$) there is
the equality $k_1=k_2=1.8m_\pi$.
It is $\omega_c$ which obeys the inequality Re$\,\omega^2_c\leq0$.
All branches depend on the isobar width: at
$\Gamma_\Delta=0$ the branch $\omega_c(k)$ is pure imaginary and
$\omega^2_c\leq 0$.  Recall that $\rho_0$ is the equilibrium density
of the matter, $\rho_0=2p^3_{F0}/3\pi^2$, $p_{F0}=268$ MeV.

The paper is organized as follows. In section 2 the
 particle-hole polarization operator $\Pi$ is
considered in the complex $\omega$-plane.  Then the branches of
solutions of the pion dispersion equation (1) are presented on
the physical and unphysical sheets.
It is shown how the branches of solutions go across the
logarithmic cuts in the complex plane. The condensate branch
$\omega_c$ is shown in details at different densities $\rho$
and isobar width $\Gamma_\Delta$.

\section{Polarization operator and its singularities}
Here we write down the formulae for the polarization operator
used. Our expressions, as is shown below, differ in
some points from the well-known expressions \cite{1,2}.
Only $S$- and $P$-waves of the $\pi NN$ ($\pi
N\Delta$) interactions are taken into account. In this case $\Pi$ is
the sum of scalar, $\Pi_S$, and vector, $\Pi_P$, terms:
\beq
\Pi(\omega,k)\ =\ \Pi_S(\omega,k)+\Pi_P(\omega,k)\ .
\eeq

The scalar polarization operator $\Pi_S$ is constructed
using  linear PCAC equation obtained by
Gell-Mann--Oakes--Renner (GMOR) \cite{10} for the pion
mass squared in the matter:
\beq
m^{*2}_\pi\ =\
-\frac{\langle NM|\bar qq|NM\rangle(m_u+m_d)}{ 2f^{*2}_\pi}\ .
\eeq
The value $\kappa=\langle NM|\bar qq|NM\rangle$ is the scalar
quark condensate calculated in nuclear matter \cite{5};
$m_u,m_d$ are masses of the current $u$- and $d$-quarks;  $f^*_\pi$
is the pion decay constant in medium. Here we put $f^*_\pi=f_\pi=92$
MeV (more details can be found in \cite{5}).

The scalar quark condensate $\kappa$ can be expanded in a power
series with respect to $\rho$ \cite{5,11}
\beq
\kappa\ =\ \kappa_0+\rho\langle N|\bar qq|N\rangle +
terms\, with\, higher\, degrees\, of\, \rho ,
\eeq
where $\kappa_0$ is the value of  scalar quark condensate
in vacuum, $\kappa_0=-0.03$ GeV$^3$; $\langle N|\bar qq|N\rangle$
is the matrix element for scalar quark condensate in
nucleon, $\langle N|\bar qq|N\rangle\simeq8$. In this place it is
enough to keep the first two terms in (4), this correspond to
the gas approximation for $\kappa$ \cite{11}. Then we get from (3)
\beq
m^{*2}_\pi\ =\ m^2_\pi-\rho\,\frac{\langle N|\bar
qq|N\rangle (m_u+m_d)}{2f^2_\pi}\ .
\eeq
On the other hand, in the dispersion equation (1)
$m^{*2}_\pi$ is defined as
\beq
m^{*2}_\pi\ =\ m^2_\pi+\Pi(\omega,k=0)\ .
\eeq
Whilst $\Pi_P(k=0)=0$, we get from (5) and (6) the
following form for $\Pi_S$:
\beq
\Pi_S\ =\ -\rho\
\frac{\langle N|\bar qq|N\rangle(m_u+m_d)}{ 2f^2_\pi}\ .
\eeq

The P-wave polarization operator $\Pi_P$ can be written
following the papers \cite{1,2,12,13}
as a sum of nucleon and isobar polarization operators
\beq
\Pi_P\ =\ \Pi_N+\Pi_\Delta\ .
\eeq
Here $\Pi_N(\Pi_\Delta$) is equal to the sum
 of the nucleon-hole and isobar-hole
loops without  pion in the intermediate states
\cite{1,2,12}:
\beq \Pi_N\ =\ \Pi^0_N\
\frac{1+(\gamma_\Delta-\gamma_{\Delta\Delta})\
\Pi^0_\Delta/k^2}E\ ,\
\Pi_\Delta\ =\ \Pi^0_\Delta\
\frac{1+(\gamma_\Delta-\gamma_{NN})\ \Pi^0_N/k^2}E\ ,
\eeq
$$
E\ =\ 1-\gamma_{NN}\frac{\Pi^0_N}{k^2}-\gamma_{\Delta\Delta}
\frac{\Pi^0_\Delta}{k^2}+\left(\gamma_{NN}\gamma_{\Delta\Delta}
-\gamma^2_\Delta\right)\frac{\Pi^0_N\Pi^0_\Delta}{k^4}\ .
$$
The expressions for the nucleon-hole loop , $\Pi^0_N$, is
as follows:
\beq
\Pi^0_N(\omega,k)={\rm Sp}\int\frac{d^3p}{(2\pi)^3}
\Gamma^2_{\pi NN}\left[\frac{\theta(p-p_F)\theta(p_F-|\vec
p+\vec k|)}{E_{\vec p+\vec k}-E_p -\omega}+
\frac{\theta(p_F-p)
\theta(|\vec p+\vec k|-p_F)}{E_p-E_{\vec p+\vec k}+\omega}
\right].
\eeq
Here $E_p=p^2/2m^*$. The expression for $\Pi^0_\Delta$ is
analogous.

The $\pi NB$ vertex $\Gamma_{\pi NB}$ with $B$ labelling
 nucleon or $\Delta$-isobar is
\beq \Gamma_{\pi NB}\ =\
\Gamma^0_{\pi NB} d_B(k),
\eeq
\beq \Gamma^0_{\pi NN}\ =\ i\
\frac{g_A}{\sqrt2\ f_\pi}\ \chi^*(\vec\sigma\vec k)\chi,\
 \Gamma^0_{\pi N\Delta}\ =\ f_{\Delta/N}\ i\ \frac{g_A}{\sqrt2\
f_\pi}\ \chi^{*\alpha}(\vec S^+_\alpha\vec k)\chi\ ,
\eeq
where
$\chi$ is nucleon 2-spinors, {$\vec \sigma$} is nucleon
spin, $\vec S^+$ turns the spin $3/2$ into $1/2$.
In order to take into account the
non-zero baryon size, the vertex $\Gamma^0_{\pi NB}$ is multiplied
by the form factor $d_B(k)$ taken in the form $
d_B=(1-m^2_\pi/\Lambda^2_B)/(1+k^2/\Lambda^2_B)$.

The constants
$\gamma_{NN},\gamma_\Delta, \gamma_{\Delta\Delta}$
are
$$
 \gamma_{NN}\ =\
C_0g'_{NN}\left(\frac{\sqrt2\ f_\pi}{g_A} \right)^2\ ,
\gamma_\Delta\ =\ \frac{C_0g'_{N\Delta}}{f_{\Delta/N}}
\left(\frac{\sqrt2\ f_\pi}{g_A}\right)^2\ ,
\gamma_{\Delta\Delta}\ =\ \frac{C_0g'_{\Delta\Delta}}{
f^2_{\Delta/N}}\left(\frac{\sqrt2\ f_\pi}{g_A}\right)^2\ ,
$$
where $C_0$ is the normalization factor ($C_0=\pi^2/(p_Fm^*)$)
and $g's$ are the constants of the effective quasi-particle--quasi-hole
interaction in nuclear matter \cite{1,2}.
The calculation results are given for the following
set of parameters \cite{1,2,12} :
\beq
f_{\Delta/N}\simeq2,\ \Lambda_N=0.667 GeV,\ \Lambda_\Delta=1 GeV,\
 g_A=1, \ f_\pi=92 MeV,
\eeq
$$
g'_{NN}=1.0, \ g'_{N\Delta}=0.2, \ g'_{\Delta\Delta}=0.8.
$$
When the parameters $f_{\Delta/N}, g', \Lambda_B$ vary within the
experimentally acceptable limits the dispersion equation solutions change
quantitatively but not qualitatively.

Later on, we can perform integration in (10) in two ways.

1. To integrate separately the first and the second terms.
In this way there appears the expression for $\Pi^0_N$ as follows:
\begin{eqnarray}
\Pi^0_N(\omega,k) &=&
-4\left(\frac{g_A}{\sqrt2\,f_\pi}\right)^2k^2\left[\Phi_N(
\omega,k)+\Phi_N(-\omega,k)\right]d^2_N(k)\ ,\\
\Phi_N(\omega,k) &=& \frac{m^*}k\ \frac1{4\pi^2}\left(
\frac{-\omega m^*+kp_F}2-\omega m^*\ln\left(\frac{ \omega
m^*}{\omega m^*-kp_F+k^2/2}\right)\ +  \right.\nonumber\\
&+&\left. \frac{(kp_F)^2-(\omega m^*-k^2/2)^2}{2k^2}\ln\left(
\frac{\omega m^*-kp_F-k^2/2}{\omega
m^*-kp_F+k^2/2}\right)\right)
\end{eqnarray}
at $0\le k\le2p_F$.  At $k\ge2p_F$ $\Phi_N(\omega,k)$ is
 Migdal's function:
\beq \Phi_N(\omega,k) =\ \frac1{4\pi^2}\
\frac{m^{*3}}{k^3}
\left[\frac{a^2-b^2}2\ln\left(\frac{a+b}{a-b}\right)-ab\right]\,
\eeq
where $a=\omega-(k^2/2m^*)$, $b=kp_F/m^*$.
Consider now which cuts has the polarization
operator $\Pi^0_N(\omega,k)$ (14)--(16)  in the
$\omega$-plane.  We can  see that at $k\le2p_F$ there are two cuts
(define them as $I$ and $II$). They are related to
the first and second logarithms in (15).  The cuts are situated
within the intervals:
\beq
 I:\quad 0\ \le\ \omega\ \le\ \frac{kp_F}{m^*} -\frac{k^2}{2m^*} ,\,
\qquad II:\quad \frac{kp_F}{m^*}-\frac{k^2}{2m^*}\ \le\ \omega\ \le\
\frac{kp_F}{m^*}+\frac{k^2}{2m^*}\ .
\eeq
Since  $\Pi^0_N$ is symmetrical under the replacement
$\omega \leftrightarrow-\omega$, the cuts of
$\Phi_N(-\omega,k)$ are placed symmetrically on the negative
semiaxis. Thus  $\Pi^0_N$ has four cuts in the complex
$\omega$-plane, they  are shown in Fig.1.

2. The other way of integration in (10) gives a well-known
expression of $\Pi^0_N$ through  Migdal's functions.
To follow it, let us do a substitution in (10)
$\theta(p-p_F)\rightarrow 1-\theta(p_F-p)=1-n(p)$. Then, after the
integration, $\Phi_N(\omega,k)$ takes a well-known form (16) for
all values of $k$ \cite{1,2,12}.  Undoubtedly, the expression (14) is
the same for both integration ways. Now $\Pi^0_N$ has not four
cuts in $\omega$-plane but two (overlapping).  It can be seen from
the expressions (10), (15), that the cut $I$ is the sum of two
overlapping cuts.  The search for dispersion equation solutions
 becomes more convenient and obvious when we work with one cut but
not with two overlapping ones. It is difficult to follow the solution
in the case 2, therefore we use equations (14)--(16) for $\Pi^0_N$.

Now let us turn to the polarization operator $\Pi^0_\Delta$,
which is the isobar--nucleon-hole loop. Since the isobar
Fermi surface is absent at the nuclear densities and isobar
momentum is unrestricted,  there is no problem
 discussed above and $\Pi^0_\Delta(\omega,k)$ reads:
\beq \Pi^0_\Delta=\
-\frac{16}9\left(\frac{g_A}{\sqrt2\,f_\pi}\right)^2
f^2_{\Delta/N}k^2\bigg[\Phi_\Delta(\omega,k)+\Phi_\Delta(
-\omega,k)\bigg]d^2_\Delta(k)\ .
\eeq
The function $\Phi_\Delta(\omega,k)$ is expressed through  Migdal's
functions (16) with $a=\omega-(k^2/2m^*)-\Delta m$, $b=kp_F/m^*$.
 The mass difference, $\Delta m=m_\Delta-m$, is the
following: Re$(\Delta m)=292$ MeV and Im$(\Delta
m)=-\Gamma_\Delta/2$.  The cuts of $\Phi^0_\Delta(\omega,k)$ are
shown in Fig.1.  At $\omega\ge0$ the cut  is in the interval
\beq
\frac{k^2}{2m^*}+\Delta m-\frac{kp_F}{m^*}\ \le\ \omega\ \le\
\frac{k^2}{2m^*}+\Delta m+\frac{kp_F}{m^*}\ .  \eeq The cut is
shifted into the complex plane in the value $-i\Gamma_\Delta/2$.

In this paper Landau equation \cite{6} is used for the
nucleon effective mass
\beq
m^*\ =\ \frac m{1+(2mp_F/\pi^2)f_1}\ .
\eeq
Unknown parameter $f_1$ is fixed by the condition
$m^*(p_F=p_{F0})=0.8m$.

\section{Solutions of the dispersion equation}
In this section the solutions of the dispersion equation (1) are
presented. The solution branch $\omega_c$
emerges on the physical sheet of the complex $\omega$-plane at
$p_F=283$ MeV for the parameter values (13).
 The figures for the zero sound branch
$\omega_s(k)$,  pion branch $\omega_\pi(k)$ and isobar
branch $\omega_\Delta(k)$ are presented for $\Gamma_\Delta=115$
MeV at $p_F=$268 and 290 MeV (i.e. at the equilibrium density
and at density slightly larger than critical one).  Some special
cases, with the other values of $p_F$ and $\Gamma_\Delta$, are
considered as well.

\vspace{0.5cm}
{\bf Branch $\omega_s(k)$.} (Fig.2a) The branch $\omega_s(k)$ is shown
for $p_F=268$ and $290$ MeV (curves 1 and 2 correspondingly).  The
branch begins at $\omega_s(k=0)=0$, then while $k$ increases, moves
practically along the real axis.  At $k_f=0.430m_\pi$ for $p_F=290$ MeV
(at $k_f=0.436m_\pi$ for $p_F=268$ MeV) goes under the cut $II$ (17),
this corresponds to the decay of $\omega_s$ into real nucleon and
the hole.

\vspace{0.5cm}
{\bf Branch $\omega_\Delta(k)$.} (Fig.2b)
The isobar branch $\omega_\Delta(k)$ begins at
$\omega=\Delta m$ at $k=0$ and ends on the isobar cut (19) at
$k_f=5.1m_\pi$ for $p_F=268$ MeV ($k_f=4.8m_\pi$ for $p_F=290$ MeV).

\vspace{0.5cm}
{\bf Branch $\omega_\pi(k)$.} (Fig.2c)
The pion branch starts at $k=0$ in $\omega_\pi=m^*_\pi$ (see
(5),(6)). The beginning of $\omega_\pi(k=0)$ is shifted to the
smaller then $m_\pi$ values when GMOR \cite{10} is used to
determine $m^*_\pi$.
The pion branch ends on  physical sheet under
the isobar cut, this corresponds to the decay of pion into
isobar and  nucleon hole. It takes place at $k_f=3.5m_\pi$ for
$p_F=268$ MeV $(k_f=3.8m_\pi$, $p_F=290$ MeV).

\vspace{0.5cm}
{\bf Branch $\omega_c(k)$.} (Fig.2d, 3a,b)
While the density increases one more branch  of solutions,
$\omega_c(k)$,  appears on the physical sheet. It emerges at
$p_F\ge283$ MeV when parameter values (13) are used. In Fig.2d the pion
branch $\omega_\pi(k)$ and condensate branch $\omega_c(k)$ are
presented at $p_F=290$ MeV.  The dashed piece of  $\omega_c(k)$
belongs to the upper unphysical sheet of the logarithmic cut $I$ (17)
(Fig.1). The branch $\omega_c(k)$ starts at $k=0$ at the same point as
$\omega_\pi(k)$ and moves onto the unphysical sheet; at
$k=k_1=1.3m_\pi$ the branch goes down to the physical
sheet and at $k=k_2=2.3m_\pi$ moves back to the same
unphysical sheet. In the momentum interval $(k_1,k_2)$  one can
follow over all the branches shown in Fig.2 and check that $\omega_c$
does not belong to any branches studied before ($\omega_s, \omega_\pi,
\omega_\Delta$).

The branch $\omega_c$ depends on the isobar width
$\Gamma_\Delta$. Decreasing the isobar width we see that the
isobar cuts move to the real axis and $\omega_s$, $\omega_\pi$
and $\omega_\Delta$ have  smaller imaginary parts. When
$\Gamma_\Delta=0$ the branches $\omega_s$, $\omega_\pi$ and
$\omega_\Delta$ are real. On the contrary, $\omega_c(k)$
moves to the imaginary axis with decreasing  $\Gamma_\Delta$
(Fig.3a). At $\Gamma_\Delta=0$ we have  pure imaginary solutions
on the physical sheet: $\omega^2_c\leq0$. 

In Fig.3b the branch $\omega_c(k)$ is shown at the different
densities: $p_F=280,290,300,360$ MeV.
When $p_F=280$ MeV the whole
 branch (curve 1) is located on  unphysical sheet.
At $p_F=283$ MeV the branch touches the real axis (not shown).
For $p_F>283$ Mev the part of $\omega_c(k)$ in the interval
$(k_1,k_2)$ is placed on the physical sheet.
At the critical density
$\rho=rho_c$ and $\Gamma_\Delta=0$ $\omega_c(k)$ touches the 
real axis at the point $\omega_c(k)=0.$ 

It was shown in paper \cite{5} that the appearance of $\omega_c$
on the physical sheet results not only in  pion condensation
but in restoration of chiral symmetry in the nuclear
matter at critical density as well.

\section{Conclusion}
In the paper the solutions of pion
dispersion equation are considered in details in the complex
$\omega$-plane.  It is shown that, besides the well-known
solutions with quantum numbers $0^-$ (zero spin-isospin
sound, pion and isobar waves), there exists the fourth branch
$\omega_c(k)$.  It is the branch which obeys the condition
$\omega^2_c\leq0$, therefore it is responsible for instability
of the ground state.  Such instability can indicate
 the beginning of 'pion condensation'. We demonstrate that at
the density less than critical one, $\rho<\rho_c$ the branch
$\omega_c(k)$ is situated on  unphysical sheet and at
$\rho\ge\rho_c$ it comes on physical one.

\subsection*{Acknowledgments} I am grateful to M.G. Ryskin
for the important and fruitful discussions during the work.
I thanks E.G. Drukarev and E.E. Saperstein for useful
discussions. This work was supported by RFFI grant 96-15-96764.

\section{Figure captions}
Fig.1.
The cuts on the physical sheet of the complex
$\omega$-plane of polarization operators $\Pi^0_N$,
$\Pi^0_\Delta$, corresponding to equations
(14), (17), (18), (19). The cuts are presented at $p_F=290$ MeV,
$k=m_\pi$.\\
Fig.2.
Branches of solutions of (1) in the complex $\omega$-plane.
Curves 1 and 2 stand for $p_F=268, 290$ MeV, correspondingly.
The dashed pieces of curves are situated on  unphysical sheets.
a) Zero spin-isospin sound branch $\omega_s(k)$. The curves are
presented up to 1.6$m_\pi$.
b) The isobar branch $\omega_\Delta$. The horizontal dashed line
is a logarithmic cut (19) for $p_F=290$ MeV at momentum
$k$ when $\omega_\Delta(k)$ is on the cut.
c) The pion branch $\omega_\pi$. The horizontal dashed line is a
logarithmic cut (19) for $p_F=290$ MeV at momentum
$k$ when $\omega_\pi$ is on the cut.  d) The total picture at
$p_F=290$ MeV for pion branch $\omega_\pi(k)$ and condensate branch
$\omega_c$(k).\\
Fig.3.  Condensate branch $\omega_c$
in the complex $\omega$-plane.
Here the dashed pieces of branches
are on the physical sheet, but  solid lines belong to unphysical
sheet.
a) The branch $\omega_c$ is
presented at $p_F=290$ MeV for different values of isobar width:
$\Gamma_\Delta= 0,10,50,115$ MeV (curves 1,2,3,4,
correspondingly); $\omega_c(k=0)=0.744m_\pi$.
b) The branch $\omega_c$ is presented at
$\Gamma_\Delta=115$ MeV for different values of Fermi momenta
$p_F=280, 290, 300, 360$ MeV (curves 1,2,3,4, correspondingly).
For $p_F$=300 and 360 MeV only the part of the branch, which
is placed on the physical sheet, is shown (the curves 3 and 4).
The whole branch for $p_F=280$ MeV (curve 1) is on unphysical sheet.

\newpage

\begin{figure}
\vspace {-9.0cm}
\centerline{\epsfig{file=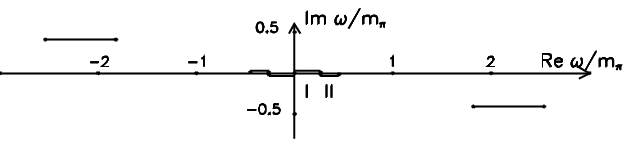,width=14cm}}
\caption{}
\end{figure}

\newpage

\begin{figure}
\centerline{\epsfig{file=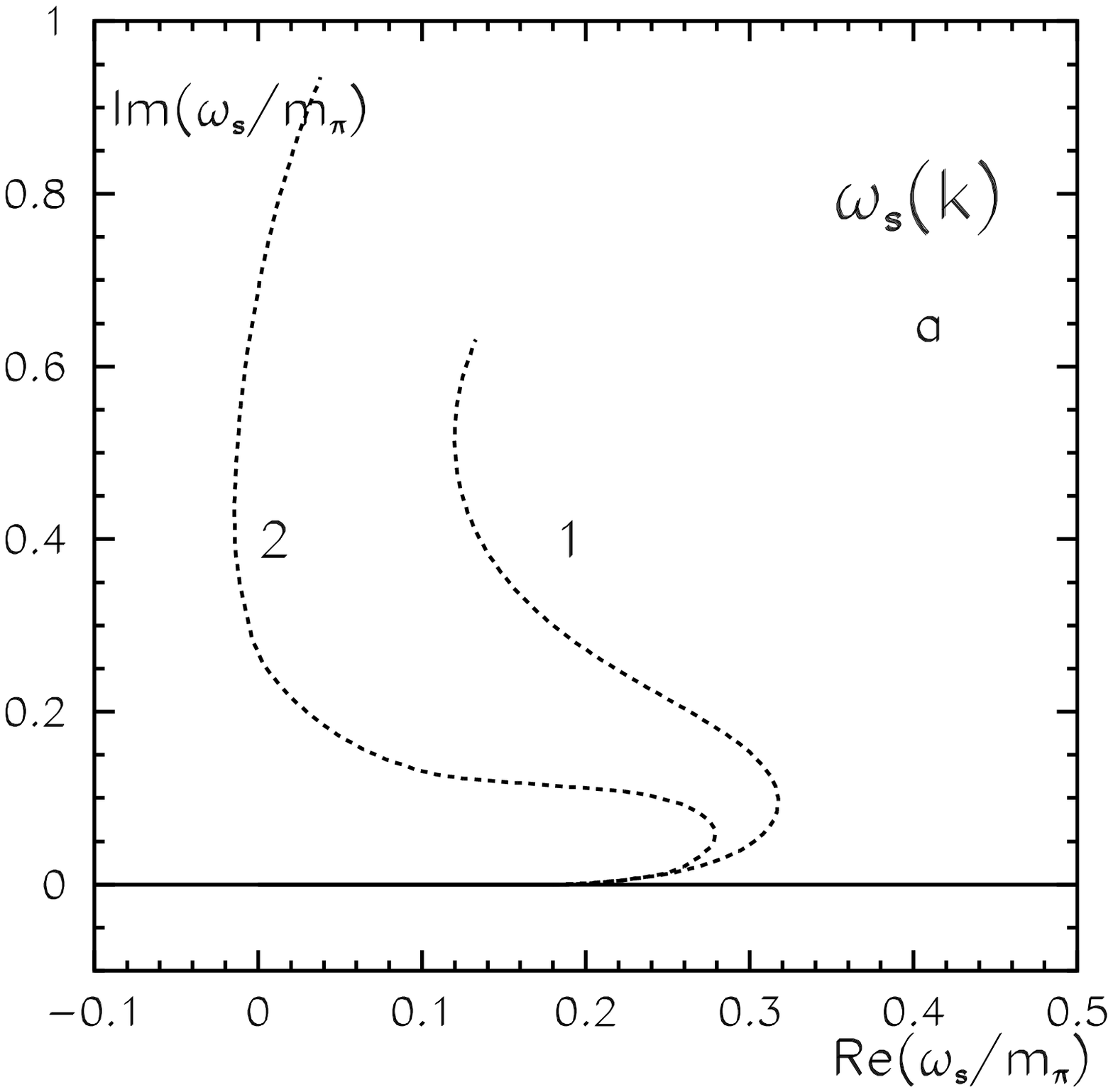,width=8cm}
\epsfig{file=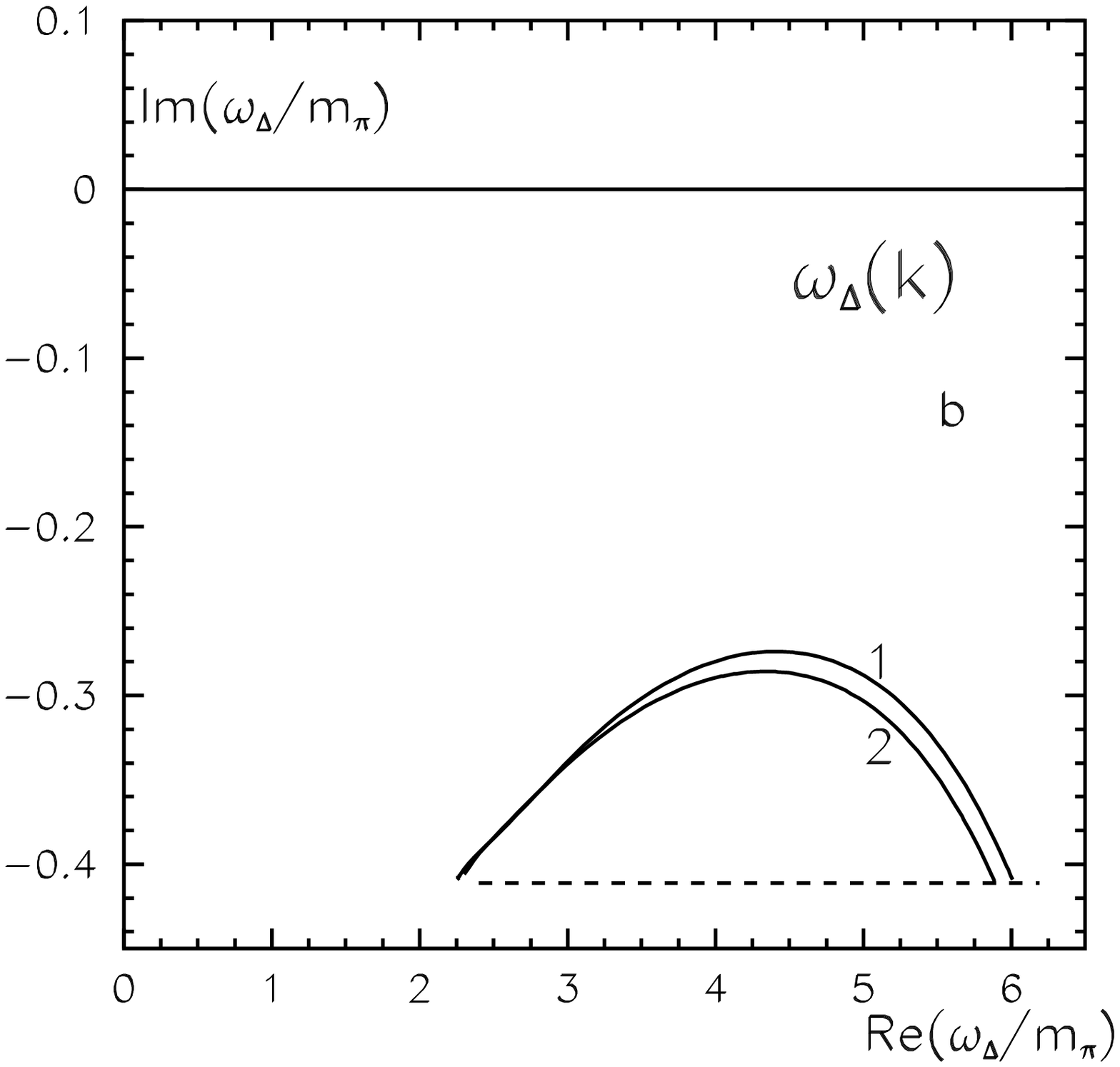,width=8cm}}
\centerline{\epsfig{file=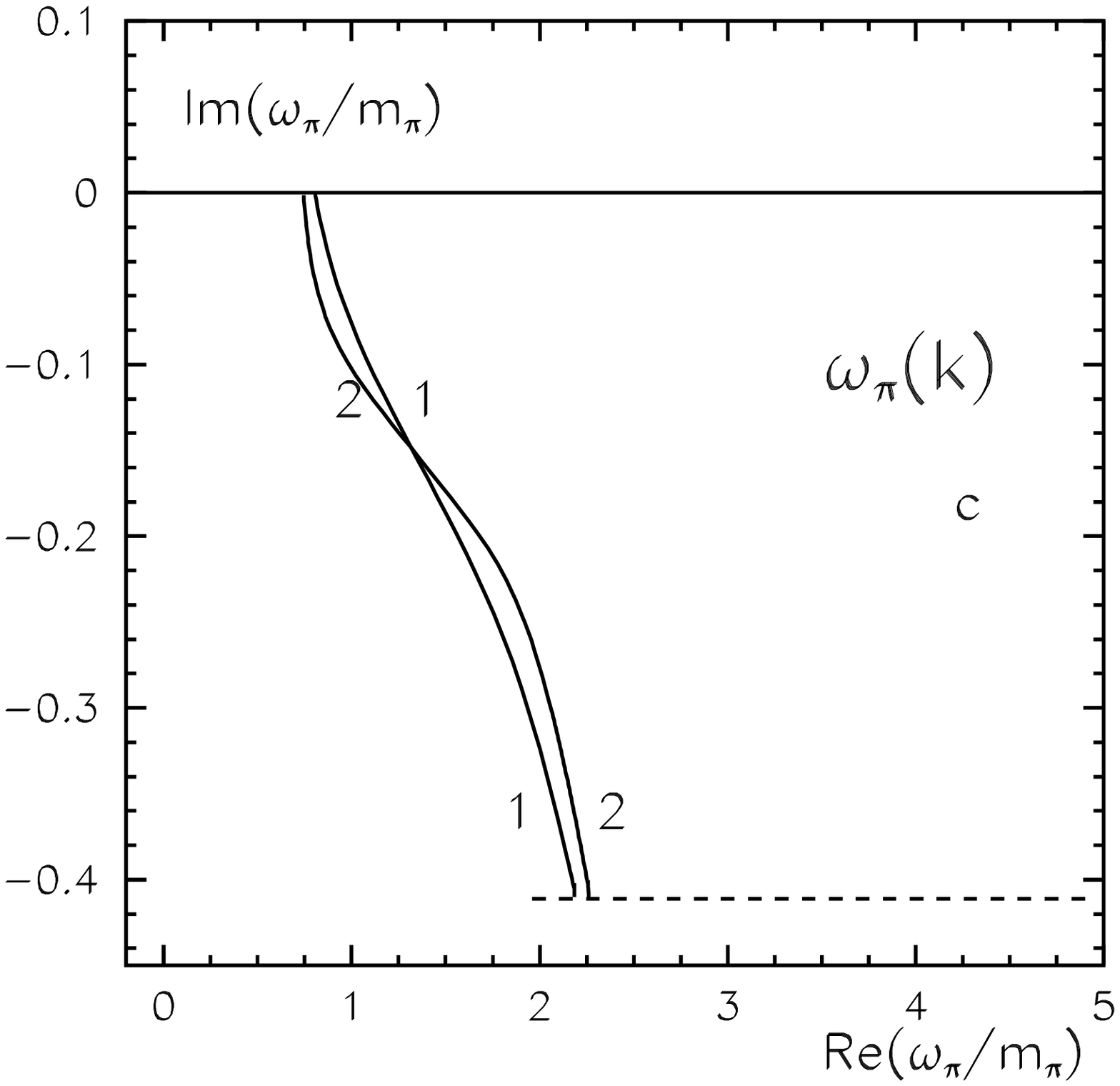,width=8cm}
\epsfig{file=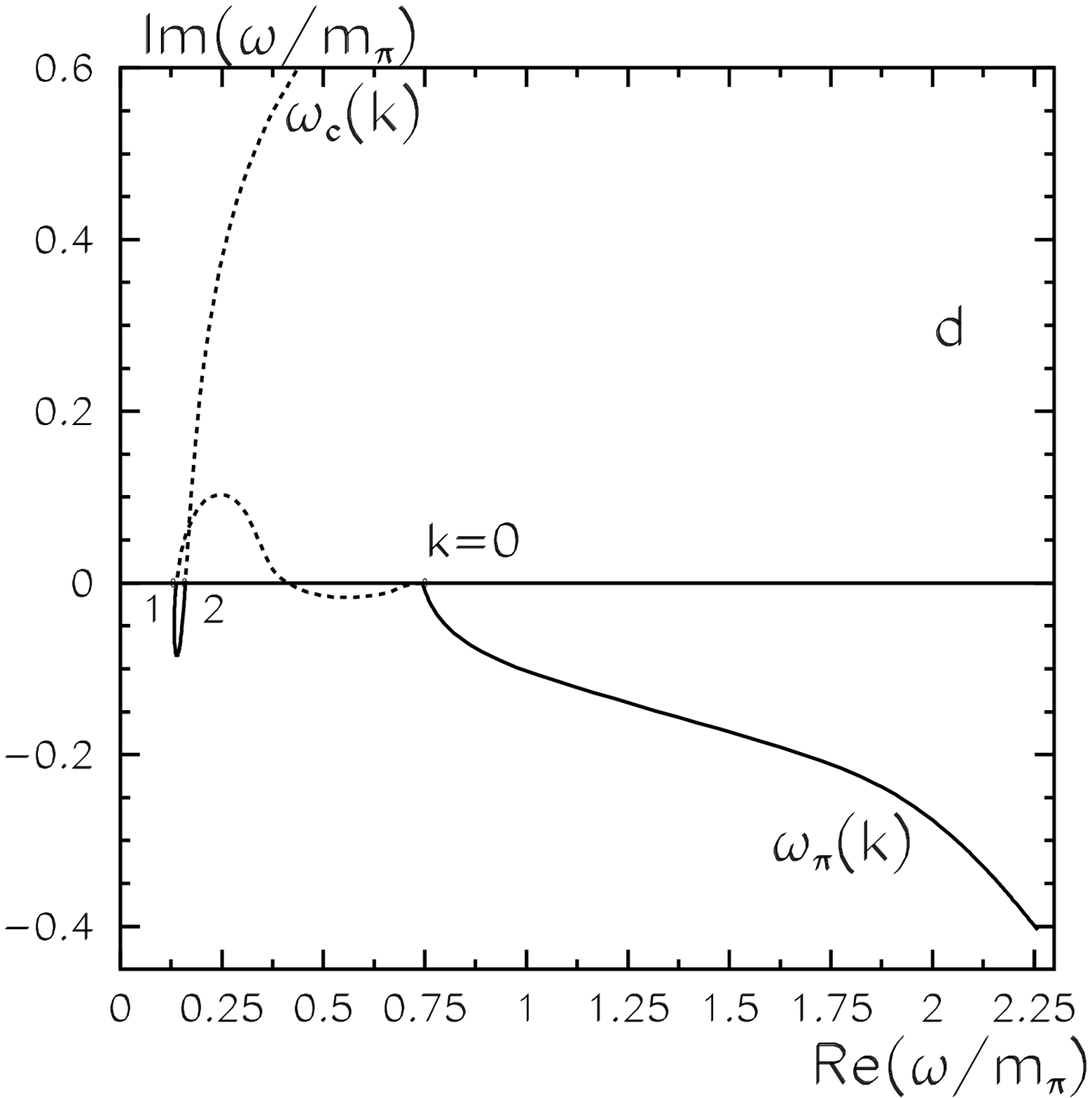,width=8cm}}
\caption{}
\end{figure}

\begin{figure}
\centerline{\epsfig{file=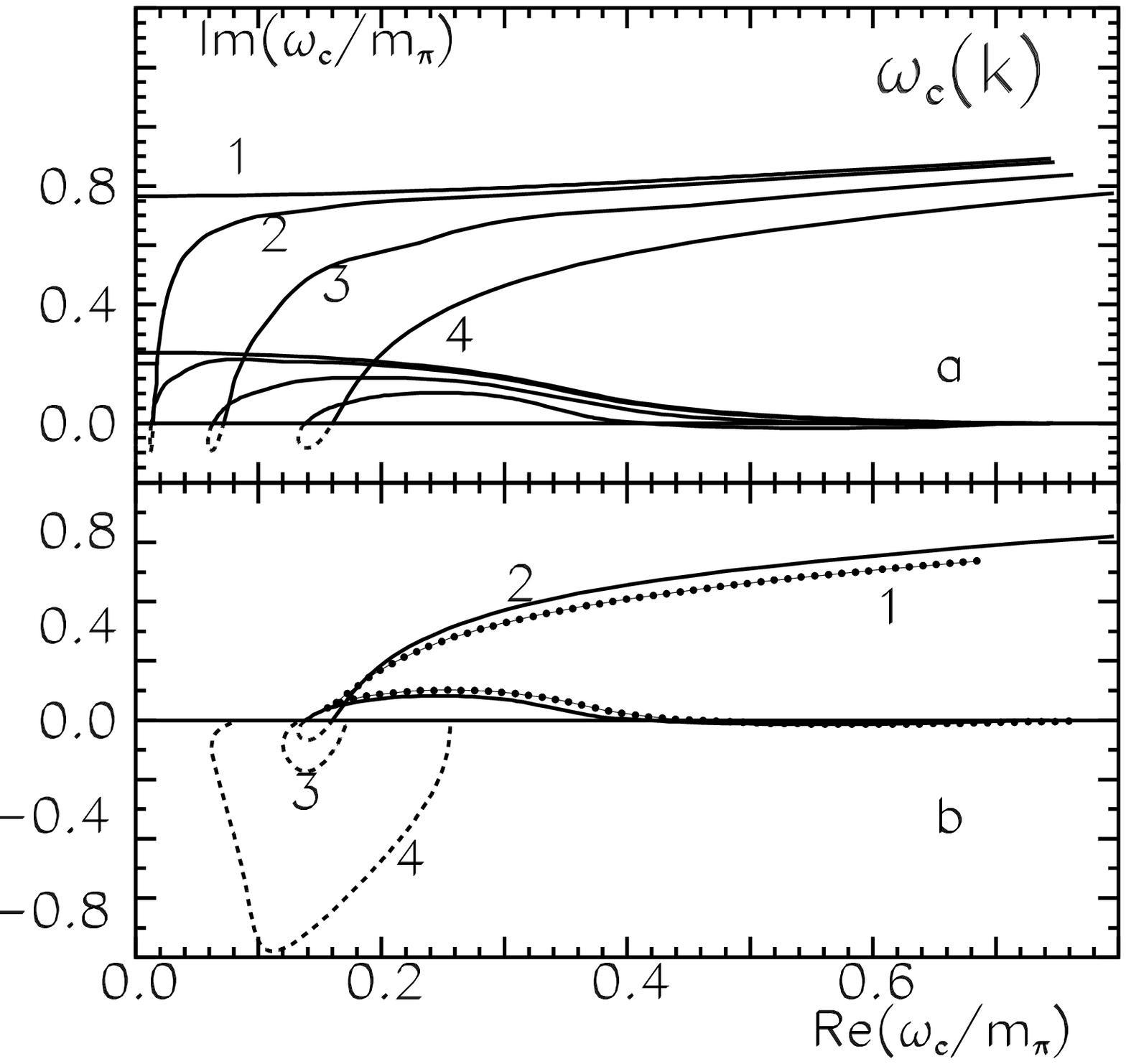,width=9cm}}
\caption{}
\end{figure}

\end{document}